\documentclass[aps,nofootinbib,prd,superscriptaddress,tightenlines,notitlepage,twocolumn,showpacs,floatfix]{revtex4-1}
\usepackage[colorlinks]{hyperref}
\usepackage{tabularx}
\usepackage{graphicx}
\usepackage{subfigure}
\usepackage{amssymb,bm,tensor}
\usepackage{textcomp}
\usepackage{xcolor}
\usepackage{amsmath}
\usepackage[varg]{txfonts}
\usepackage{enumerate}
\usepackage{mathtools}
\usepackage[normalem]{ulem}
\usepackage{bm} 
\usepackage[OT1]{fontenc}
\usepackage{aas_macros}
\usepackage{verbatim}

\newcommand{\PHY}{\affiliation{School of Physics and State Key Laboratory of
Nuclear Physics and Technology, Peking University, Beijing 100871, China}}
\newcommand{\MPIfR}{\affiliation{Max-Planck-Institut f\"ur Radioastronomie, Auf
dem H\"ugel 69, D-53121 Bonn, Germany}}
\newcommand{\KIAA}{\affiliation{Kavli Institute for Astronomy and Astrophysics,
Peking University, Beijing 100871, China}}
\newcommand{\CICQM}{\affiliation{Collaborative Innovation Center of Quantum
Matter, Beijing, China}}
\newcommand{\CHEP}{\affiliation{Center for High Energy Physics, Peking
University, Beijing 100871, China}}

\begin{document}

\title{Bounding the mass of graviton in a dynamic regime with binary pulsars}
\date{\today}
\author{Xueli Miao}\PHY
\author{Lijing Shao}\email{lshao@pku.edu.cn}\KIAA\MPIfR
\author{Bo-Qiang Ma}\email{mabq@pku.edu.cn}\PHY\CICQM\CHEP

\begin{abstract}
In Einstein's general relativity, gravity is mediated by a massless spin-2
metric field, and its extension to include a mass for the graviton has profound
implication for gravitation and  cosmology.  In 2002, \citet{Finn:2001qi} used
the gravitational-wave (GW) back-reaction in binary pulsars, and provided the
first bound on the mass of graviton. Here we provide an improved analysis using
9 well-timed binary pulsars with a phenomenological treatment.  First,
individual mass bounds from each pulsar are obtained in the frequentist
approach with the help of an ordering principle.  The best  upper limit on the
graviton mass, $m_{g}<3.5\times10^{-20} \, {\rm eV}/c^{2}$ (90\% C.L.), comes
from the Hulse-Taylor pulsar PSR~B1913+16.  Then, we combine individual pulsars
using the Bayesian theorem, and get $m_{g}<5.2\times10^{-21} \, {\rm eV}/c^{2}$
(90\% C.L.) with a uniform prior for $\ln m_g$. This limit improves the
Finn-Sutton limit by a factor of more than 10. Though it is not as tight as
those from GWs and the Solar System, it provides an independent and
complementary bound from a {\it dynamic regime}.
\end{abstract}

\maketitle


\section{Introduction}
\label{sec:intro}

In the standard model of particle physics, forces are mediated by gauge bosons:
photons for the electromagnetic force, gluons for the strong force, $W^\pm$ and
$Z$ bosons for the weak force. While gravity has not been unified with the
other three interactions, it is widely believed that gravitation is mediated by
a massless spin-2 graviton, at least in Einstein's theory of general relativity
(GR) \cite{Weinberg:1972kfs}. Strictly speaking, the existence of graviton is
not experimentally confirmed yet \cite{Patrignani:2016xqp}; see {\it e.g.}
\citet{DYSON:2013jra}.

The masslessness of graviton was challenged by \citet{Fierz:1939ix} back in
1939. Later it was found that their approach introduces a discontinuity, the
so-called vDVZ discontinuity. Because of the vDVZ discontinuity, when the mass
of graviton, $m_g$, goes to zero, GR cannot be recovered~\cite{Iwasaki:1971uz,
vanDam:1970vg, Zakharov:1970cc}. This problem is resolved when the so-called
screening mechanisms take place ({\it e.g.} the Vainshtein
mechanism~\cite{Vainshtein:1972sx}). However, a finite-range gravity has
Boulware-Deser ghosts \cite{Boulware:1973my}, that are fortunately tamed by the
recent ghost-free de Rham-Gabadadze-Tolley (dRGT) gravity model
\cite{deRham:2010kj, deRham:2014zqa}. With all the pathologies cured
theoretically, it is intriguing to look for the clues of the massive graviton
in experiments and observations.

At present, there are many ways to study the graviton mass, and some of them
give very tight bounds~\cite{deRham:2016nuf}.
\begin{itemize}
    \item Using the modified dispersion relation in the propagation of
      gravitational waves (GWs)~\cite{Will:1997bb}, the LIGO/Virgo
      Collaboration constrained the graviton mass to be,
      \begin{align}
        m_g \leq 1.2\times10^{-22} \, {\rm eV}/c^2 \quad \mbox{(90\% C.L.)}\,,
      \end{align}
    with the first event GW150914~\cite{TheLIGOScientific:2016src}. The latest
    mass bound comes from the binary black hole signals by combining the
    LIGO/Virgo catalog GWTC-1~\cite{LIGOScientific:2019fpa},
    \begin{align}
	m_g \leq 5.0 \times 10^{-23} \, {\rm eV}/c^2 \quad \mbox{(90\% C.L.)}\,.
    \end{align}
    \item In the Solar System, to leading-order approximation the perihelion
      advance correction per orbit induced by the Yukawa potential from massive
      graviton is~\cite{Will:2018gku},
    \begin{align}
	\Delta \varpi &= \pi \left( \frac{a}{\lambda_g} \right)^2
	\frac{1}{\sqrt{1-e^2}}\,,
    \end{align}
    where $\varpi$ is the longitude of perihelion measured from a fixed
    reference direction, $e$ is the orbital eccentricity, $a$ is the semimajor
    axis of the orbit, and $\lambda_{g}$ is the Compton wavelength of graviton.
    While the GR precession, $\Delta \varpi_{\rm GR} = 6\pi G M_\odot /
    c^2a\left( 1-e^2 \right)$, is larger for orbits closer to the Sun, the
    massive graviton effect grows with distance from the
    Sun~\cite{Will:2018gku}. The data on the perihelion advance of the Mars
    obtained from the Mars Reconaissance Orbiter lead to a credible upper bound
    on $m_g$~\cite{Will:2018gku},
    \begin{align}
      m_{g}<(6-10)\times 10^{-24}\,{\rm eV}/c^{2}\,.
    \end{align}
    \item Besides the above two examples, model-dependent limits include the
      one inferred from the stability of black-hole metric
      \cite{Brito:2013wya}, those from the galactic and cluster dynamics
      \cite{Goldhaber:1974wg}, and gravitational weak lensing
      \cite{Choudhury:2002pu}. Although they provide better results, we still
      have large uncertainties in these models, concerning, {\it e.g.} the
      distribution of the dark matter.
    \item Finn and Sutton~\cite{Finn:2001qi, Sutton:2001yj} presented a
      linearized GR augmented with a gravitational mass term in a
      phenomenological way, and used the observed orbital decay from binary
      pulsars PSRs~B1913+16 and B1534+12 to get a mass bound 
    \begin{equation}
      m_{g}<7.6\times10^{-20}\,{\rm eV}/c^{2} \quad \mbox{(90\% C.L.)}\,.
    \end{equation}
    This is the first bound which is obtained from a {\it dynamic regime} as
    opposed to the static Yukawa potential and the kinematics of GW
    propagation.  Although strictly speaking, in a Lorentz-invariant theory of
    massive graviton, the helicity-0 mode is neccessarily present and its
    interactions lead to a Vainshtein screening
    mechanism~\cite{deRham:2016nuf}, the limit obtained from
    \citet{Finn:2001qi} is {\it phenomenologically indicative} for the
    helicity-2 modes and practically useful for simple comparisons. Improved
    analysis within the cubic Galileon can be found in
    Refs.~\cite{deRham:2012fw, deRham:2012fg}. 
\end{itemize}

In this paper we use the method in Ref.~\cite{Finn:2001qi} and present an
improved analysis with binary pulsars. We use the updated data of PSRs~B1913+16
and B1534+12, and the data of another 7 carefully-chosen, well-timed binary
pulsars to give individual mass bounds with the Finn-Sutton
method~\cite{Finn:2001qi}.  We choose an ordering principle to specify uniquely
the acceptance interval, which makes sure to avoid an unphysical or an empty
confidence interval~\cite{Feldman:1997qc}. It ensures that we only deal with a
positive mass. After getting individual mass bounds, we  combine  the
individual  data  using  the Bayesian  theorem  in a coherent approach to give
a final mass bound. The combined limit, 
\begin{equation}
m_{g}<5.2\times10^{-21} \, {\rm eV}/c^{2}  \quad \mbox{(90\% C.L.)} \,,
\end{equation}
is not as tight as other limits, yet providing an independent and complementary
bound from a {\it dynamic regime}.

The paper is organised as follows. In the next section, we present the
Finn-Sutton method~\cite{Finn:2001qi, Sutton:2001yj}, and the statistical
framework to deal with truncated priors and to combine multiple independent
observations. The method is applied to 9 binary pulsars in
Sec.~\ref{sec:limit}, where improved bounds are obtained. The last section
discusses the comparison between our results and some previous ones.

In a unit system where $\hbar = c = 1$, to convert between different
quantities, it is useful to remember, $1\,{\rm eV} = 1.8 \times 10^{-36} \,{\rm
kg} = \left( 2.0 \times 10^{-7} \, {\rm m} \right)^{-1}$.

\section{Theoretical framework}

In Sec.~\ref{sec:graviton}, we review the Finn-Sutton
method~\cite{Finn:2001qi}. In Sec.~\ref{sec:stats}, in addition to the
frequentist approach, we extend the analysis with a Bayesian treatment.

\subsection{Linearized gravity with a massive graviton}
\label{sec:graviton}

We consider a phenomenological action for linearized gravity with a mass
term~\cite{Visser:1997hd, Finn:2001qi},
\begin{align}
    I =& \frac{1}{64\pi} \int {\rm d}^4 x \left[ \partial_\lambda h_{\mu\nu}
	\partial^\lambda  h^{\mu\nu} - 2 \partial^\nu h_{\mu\nu}
	\partial_\lambda h^{\mu\lambda} + 2 \partial^\nu h_{\mu\nu}
	\partial^\mu h   \right. \nonumber \\
	& \left. - \partial^\mu h \partial_\mu h - 32 \pi h_{\mu\nu}
    T^{\mu\nu} + m_g^2 \left( h_{\mu\nu} h^{\mu\nu} - \frac{1}{2} h^2
    \right)\right] \,,
    \label{eq:action}
\end{align}
where $h_{\mu\nu} \equiv g_{\mu\nu} - \eta_{\mu\nu}$ with $\left| h_{\mu\nu}
\right| \ll 1$, and $h \equiv \tensor{h}{^\mu_\mu}$.  The mass term is unique
for a linearized gravity with only the helicity-2 modes when, 
\begin{enumerate}
  \item the derived wave equation takes the standard form with an
  $h$-independent source [see Eq.~(\ref{eq:field})], and
  \item the predictions of GR are recovered when $m_g
  \to0$~\cite{Visser:1997hd, Finn:2001qi}.
\end{enumerate}
The action (\ref{eq:action}) is viewed as an effective-field-theoretic
{\it model}, instead of a full {\it theory}. When only considering the
perturbation of the metric field around a Minkowski spacetime, for a
linearized gravity the form of the action is uniquely determined if the above
two requirements are met~\cite{Finn:2001qi, Visser:1997hd}.

We are aware of the fact that, if the action were taken as {\it a full theory}
instead of {\it a phenomenological effective treatment}, some pathological
features could appear ({\it e.g.} ghosts and instabilities). A more
sophisticated theory would asks for a well-designed structure, likely equipped
with extra scalar fields in a Lorentz-invariant theory or the possibility of
Lorentz violation~\cite{deRham:2014zqa, deRham:2016nuf}.  Specifically
well-designed examples include the Dvali-Gabadadze-Porrati (DGP) model, and the
dRGT theory and its bigravity extensions~\cite{Hassan:2011zd} where the
screening mechanisms take effect.  Here in this work we try to be agnostic, and
following \citet{Finn:2001qi}, we use the action (\ref{eq:action}) for the
study.  In particular, we do not take account of the screening mechanisms nor
the propagating modes beyond helicity-2, and we work only in the linearized
Fierz-Pauli-like model. Detailed discussions on massive gravity theories can be
found in Ref.~\cite{deRham:2014zqa} and references therein.  To convert from
our phenomenologically generic limits to the mass parameter in a full theory, a
careful analysis is needed. It is worthy to note that, the well-adopted
treatment of the GW propagation~\cite{Will:1997bb, TheLIGOScientific:2016src,
LIGOScientific:2019fpa}, where the mass of graviton is fixed canonically as a
constant everywhere, is of a similar spirit, for a full massive-gravity theory
might predict a mass depending on the specific environment or epoch of the
Universe. We defer a careful analysis with specific massive gravity theories in
a future study.

Applying the Noether's theorem to the action (\ref{eq:action}), one gets an
identical effective stress-energy tensor (ESET) for GWs as that in
GR~\cite{Finn:2001qi},
\begin{align}
    \label{eq:stress}
    T_{\mu\nu}^{\rm GW} = \frac{1}{32\pi} \left\langle \partial_\mu \bar
    h_{\alpha\beta} \partial_\nu \bar h^{\alpha\beta} - \frac{1}{2}
    \partial_\mu \bar h \partial_\nu \bar h \right\rangle \,,
\end{align}
where ``$\left\langle \cdot \right\rangle$'' denotes averaging over a spatial
volume with a linear dimension larger than the wavelength of GWs, and  the
trace-reversed metric perturbation is defined by $\bar h_{\mu\nu} \equiv
h_{\mu\nu} - \frac{1}{2} \eta_{\mu\nu} h$~\cite{Finn:2001qi}.

{It is interesting to note that, the ESET (\ref{eq:stress}) is the
same as the one derived by \citet{Isi:2018miq}. They based on the Noether's
theorem in a Minkowski background with the Fierz-Pauli mass
term~\cite{Fierz:1939ix}. Although it leads to a different equation of motion
and the solutions $h_{\mu\nu}$ are different from
ours,\footnote{While the linearized Fierz-Pauli
model~\cite{Fierz:1939ix} has five independent GW polarizations, the
Finn-Sutton method~\cite{Finn:2001qi, Sutton:2001yj} concentrates solely on
the two tensor modes.} our results are applicable to the Fierz-Pauli linear
massive gravity model. In principle, concerning the nonlinearity, the
assumption about the Minkowski background is only valid outside of the
Vainshtein radius where healthy theories of massive gravity will predict the
same ESET. But in our case, binary pulsars are within the Vainshtein radius
of the Milky Way. It is still an open question whether the ESET remains the
same. One should keep the caveats when quoting our results.

Conservation of the stress tensor of matters, $\partial^\nu T_{\mu\nu} = 0$,
imposes a Lorenz-gauge-like constraint, $\partial_\nu \bar{h}^{\mu\nu} = 0$.
With this constraint, the field equation is in the standard form for a
massive particle~\cite{Weinberg:1995mt, Finn:2001qi},
\begin{align}
    \left( \square - m_g^2 \right) \bar h_{\mu\nu} + 16\pi T_{\mu\nu} = 0 \,.
    \label{eq:field}
\end{align}

Assuming slow motions for objects in a bound Keplerian orbit,
\citet{Finn:2001qi} worked out the solution of Eq.~(\ref{eq:field}) in the
frequency domain~\cite{Peters:1963ux}, and obtained the fractional corrections
to the GW radiation luminosity in GR,
\begin{align}
    \Delta \equiv \frac{L_{\rm GW} - L_{\rm GW}^{(0)}}{L_{\rm GW}^{(0)}} =
    \frac{5}{24} \frac{1}{F(e)} m_g^2 \left( \frac{c^2 P_b}{2\pi \hbar} \right)^2 \,,
    \label{eq:1}
\end{align}
where $P_b$ is the orbital period, $e$ is the orbital eccentricity, and the
superscript ``$(0)$'' denotes the corresponding quantity in GR, and $F(e)$ is a
function of the eccentricity~\cite{Finn:2001qi} (see Fig.~\ref{fig:fe}),
\begin{align}
\label{eq:Fe}
  F(e)=\frac{1+\frac{73}{24}e^{2}+\frac{37}{96}e^{4}}{\left(1-e^{2}\right)^{3}}\,.
\end{align}

It is worthy to note that, corrections to the conservative orbital dynamics can
be neglected (see footnote 1 in~\citet{Finn:2001qi}).  The standard first
post-Newtonian approximation to binary pulsars is sufficient for the
conservative dynamics~\cite{Lorimer:2005misc}.  Therefore, it complies with
pulsar observations~\cite{Wex:2014nva, Shao:2016ezh}.

\subsection{Statistical treatments}
\label{sec:stats}

Here we present the frequentist and Bayesian approaches to obtain the mass
bound of graviton, with the constraint that $m_g$ is non-negative.

\subsubsection{Frequentist confidence intervals}

Frequentist confidence intervals can be obtained by Neyman's
method~\cite{Neyman:1937uhy}, by constructing standard confidence belts.
However, there are some problems of the usual results from Neyman's
construction for lower or upper confidence limits, in particular when the
confidence interval gives an unphysical interval or an empty
set~\cite{Feldman:1997qc} (e.g., a negative mass in our cases).  Basing on the
results of the experiment to decide whether to publish an upper limit or a
central confidence interval seems to avoid the above
problem~\cite{Feldman:1997qc}.  However, the intervals obtained from this way
maybe undercover for a significant range of an unknown physical quantity.  It
means that these intervals are not confidence intervals or conservative
confidence intervals~\cite{Feldman:1997qc}.

To avoid above problems, one can choose an ordering principle which bases on
the freedom inherent in Neyman's construction to specify uniquely the
acceptance interval~\cite{Feldman:1997qc}.  This method makes intervals
automatically change from lower or upper limits to two-sided intervals.  It
avoids possible undercoverage caused by basing personal choice on the data, and
makes sure that the confidence interval is never an unphysical interval or an
empty set. Following~\citet{Finn:2001qi}, we will make use of the approach
invented by \citet{Feldman:1997qc} to bound the graviton mass from individual
pulsars.

\subsubsection{Bayesian framework}

When combining observations from multiple binary pulsars, it is convenient to
adopt the Bayesian treatment.  In Bayesian statistics, the interpretation of
probability is more general and includes the prior degree of belief, which is
updated by the data from subsequent experiments.  If the data are sufficient,
the posterior distributions are no longer dependent on the choice of prior.  We
can use observations to obtain the posterior distributions of the parameters by
the Bayesian theorem.  For our study, the posterior density function is
\cite{DelPozzo:2016ugt},
\begin{align}
    P\left( \left. m_g  \right| {\cal D}, {\cal H}, {\cal I} \right) &= \int
	\frac{P \left( \left.{\cal D}  \right| m_g, \bm{\Xi}, {\cal H}, {\cal
		I} \right) P\left( \left. m_g, \bm{\Xi} \right| {\cal H}, {\cal
		I} \right)}{P \left( \left. {\cal D} \right| {\cal H}, {\cal I}
		    \right)} {\rm d} \bm{\Xi} \,,
   \label{eq:4}
\end{align}
where ${\cal I}$ is all other relevant prior background knowledge, $\bm{\Xi}$
collectively denotes all other unknown parameters besides $m_g$, ${\cal D}$ are
data, and $\cal{H}$ is the hypothesis or the model.  In the above equation,
$P\left(m_{g},\bm{\Xi}|\cal{H},\cal{I}\right)$ is the prior probability
density, $P\left({\cal D} | m_{g},\bm{\Xi},\cal{H},\cal{I}\right)$ is the
likelihood function, and $P\left({\cal D}|\cal{H},\cal{I} \right)$ is the model
evidence which generally equals to a constant playing the role of
normalization.

\bgroup
\def\arraystretch{1.25}
\begin{table*}
    \caption{\label{tab:psr}
	Relevant parameters for the 9 binary pulsar systems that we use in this
      study. Parenthesized numbers represent the 1-$\sigma$ uncertainty in the
    last digits quoted.}
    \begin{tabular}{p{2.8cm}p{3cm}p{2.8cm}p{1.9cm}p{1.9cm}p{2.3cm}p{2.3cm}}
	\hline\hline
	PSR & $P_b \, ({\rm day})$ & $e$ & $m_p \, (M_\odot)$ & $m_c \,
	(M_\odot)$ & $\dot P_b^{\rm intr} \, (10^{-12})$ & $\Delta$ \\
	\hline
	J0348+0432 \cite{Antoniadis:2013pzd} & $0.102424062722(7)$ &
	$0.0000026(9)$ & $2.01(4)$ & $0.172(3)$ & $-0.273(45)$ &$0.05(18)$ \\
	J0737$-$3039 \cite{Kramer:2006nb, Kramer:2016kwa} & $0.10225156248(5)$ & $0.0877775(9)$
	& $1.3381(7)$ & $1.2489(7)$ & $-1.252(17)$ & $0.000(1)$ \\
	J1012+5307 \cite{Lazaridis:2009kq, Antoniadis:2016hxz} &
	$0.60467271355(3)$ & $0.0000012(3)$ & $1.83(11)$  & $0.174(7)$ &
	$-0.015(15)$ & $0.36(145)$\\
	B1534+12 \cite{Fonseca:2014qla,Stairs:2002cw} & $0.420737298879(2)$ & $0.27367752(7)$
	& $1.3330(2)$ & $1.3455(2)$ & $-0.174(11)$ & $-0.096(57)$\\
	J1713+0747 \cite{Zhu:2018etc} & $67.8251299228(5)$ & $0.0000749403(7)$
	& $1.33(10)$ & $0.290(11)$ & $0.03(15)$ & $-5000(25000)$\\
	J1738+0333 \cite{Freire:2012mg} & $0.3547907398724(13)$ &
	$0.00000034(11)$ & $1.46^{+0.06}_{-0.05}$ & $0.181^{+0.008}_{-0.007}$
	& $-0.0259(32)$ & $-0.072(130)$\\
	J1909$-$3744 \cite{Desvignes:2016yex} & $1.533449474329(13)$ &
	$0.00000021(9)$ & $1.540(27)$ & $0.2130(24)$ & $-0.006(15)$ & $2.08(521)$\\
	B1913+16 \cite{Weisberg:2016jye} & $0.322997448918(3)$ & $0.6171340(4)$
	& $1.438(1)$ & $1.390(1)$ & $-2.398(4)$ & $-0.0017(16)$\\
	J2222$-$0137 \cite{Cognard:2017xyr} & $2.44576456(13)$ &
	$0.000380940(3)$ & $1.84(6)$ & $1.323(25)$ & $-0.063(85)$ & $-1.3(117)$\\
	\hline
    \end{tabular}
\end{table*}
\egroup

\section{Constraints on the mass of graviton}
\label{sec:limit}

In this section we combine the theoretical results in Sec.~\ref{sec:graviton}
and the statistical methods in Sec.~\ref{sec:stats} to obtain bounds on the
graviton mass.

\subsection{Binary pulsars}

Timing of the periodic pulses from binary pulsars has extremely high accuracy.
A high-precision measurement usually gives a high-precision orbital period
decay, $\dot{P}_{b}$, for a relativistic binary.  It can be used to test GR
(see {\it e.g.} Ref.~\cite{Shao:2017gwu}) or to constrain other alternative
gravity theories.  Binary pulsars are strongly self-gravitating systems,
intrinsically suitable for various gravity tests~\cite{Wex:2014nva}.  When
binary pulsars radiate GWs in the inspiral process, the orbital period $P_{b}$
could have an observable change.  If the graviton mass is nonzero, the power of
gravitational emission will be different from the prediction of GR. Therefore
the orbital period decay of binary pulsars will differ from the prediction of
GR, and we can utilize $\dot{P}_{b}$ to bound the graviton mass.

For a slowly decaying Keplerian binary, the instantaneous period derivative is
proportional to the energy loss rate, $\dot P_b \propto L_{\rm
GW}$~\cite{Finn:2001qi}.  We can identify the fractional discrepancy between
the predicted decay rate and the observed decay rate with Eq.~(\ref{eq:1}), 
\begin{align}
    \frac{\dot{P}_{b}-\dot{P}^{\rm GR}_{b}}{\dot{P}^{\rm
    GR}_{b}}=\frac{L_{\rm GW}-L_{\rm GW}^{(0)}}{L_{\rm GW}^{(0)}} \equiv \Delta \,,
    \label{eq:delta}
\end{align}
where $\dot{P}^{\rm GR}_{b}$ is the value of orbital period decay in GR.
$\dot{P}^{\rm GR}_{b}$ is obtained by calculating the orbital period decay that
is caused by the emission of quadrupolar GWs~\cite{Peters:1964zz},
\begin{align}
 \dot{P}^{\rm GR}_{b}=-\frac{192\pi
 G^{5/3}}{5c^{5}}G(e)
 \left(\frac{2\pi}{P_{b}}\right)^{5/3}
 \frac{m_{p}m_{c}}{\left(m_{p}+m_{c}\right)^{1/3}}\,,
\end{align}
with $G(e)$ a function of the eccentricity (see Fig.~\ref{fig:fe}),
\begin{align}
\label{eq:Ge}
 G(e) =
 \frac{1+\frac{73}{24}e^{2}+\frac{37}{96}e^{4}}{\left(1-e^{2}\right)^{7/2}}\,.
\end{align}
Combining with Eq.~(\ref{eq:1}) and Eq.~(\ref{eq:delta}), we can loosely give
an upper limit to the squared graviton mass~\cite{Finn:2001qi},
\begin{align}
   m_{g}^{2} \lesssim
   \frac{24}{5}F(e)\left(\frac{2\pi\hbar}{c^{2}P_{b}}\right)^{2}\Delta\,.
   \label{eq:3}
\end{align}

\begin{figure}
\includegraphics[width=0.40\textwidth]{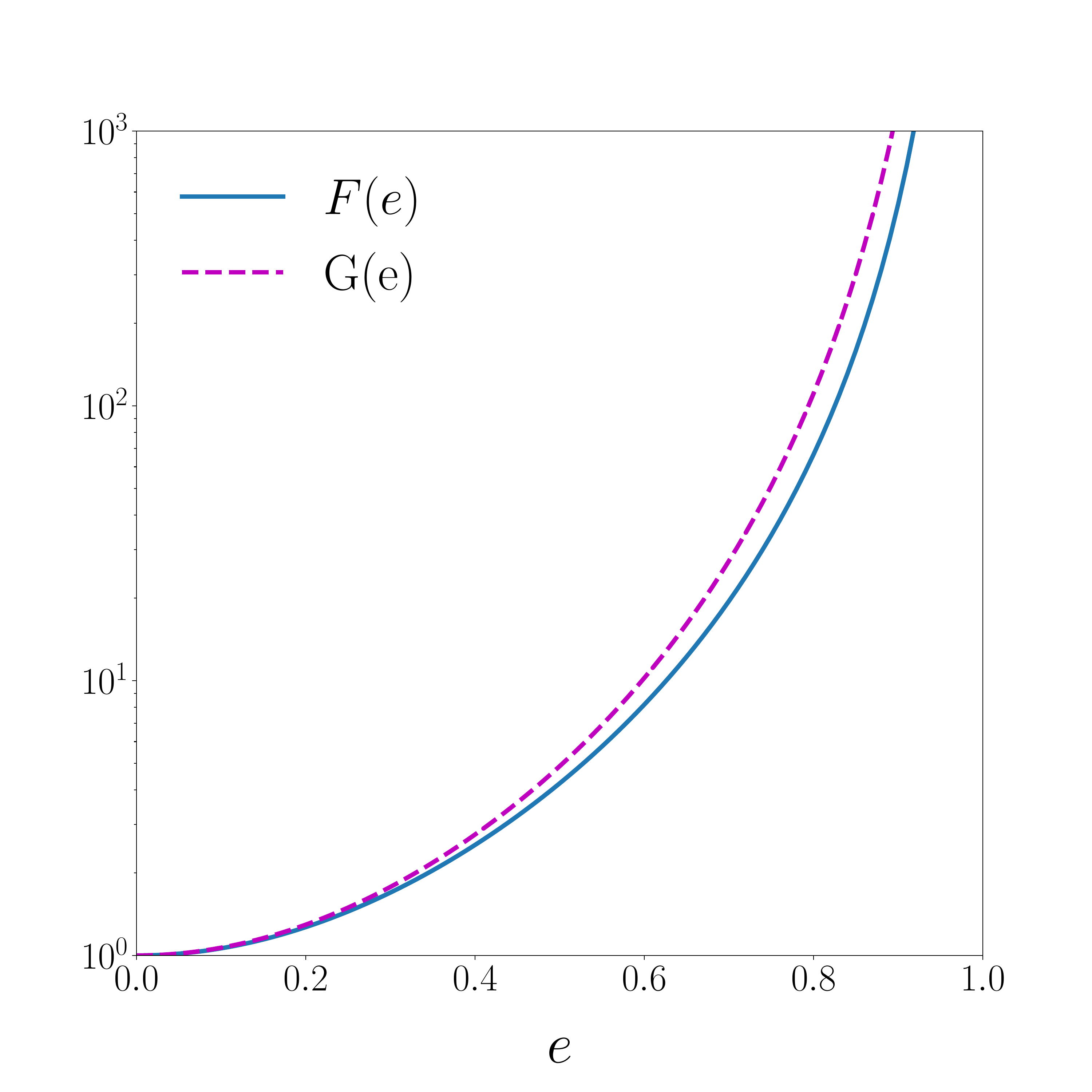}
\caption{Eccentricity functions $F(e)$ and $G(e)$, defined in Eq.~(\ref{eq:Fe})
and Eq.~(\ref{eq:Ge}), respectively. }\label{fig:fe}
\end{figure}

Quite a few factors affect the observed $\dot P_b$ (denoted as $\dot P_b^{\rm
obs}$), including the relative acceleration between the binary pulsar and the
Solar System barycenter, the kinematic effects, and so on (see
Ref.~\cite{Lorimer:2005misc} for a detailed account).  In our study the
theoretical $\dot{P}_{b}$ (including the quadrupole radation in GR and the
contribution from the massive graviton) should be compared to the intrinsic
orbital decay rate of binary pulsar systems. We should subtract other effects
from the observed orbital period
decay~\cite{1970SvA....13..562S,Damour:1990wz}, to obtain the intrinsic orbital
decay,
\begin{align}
 \label{eq:Pbdot:intr}
 \dot{P}_{b}^{\rm intr}=\dot{P}_{b}^{\rm obs}-\dot{P}_{b}^{\rm
 Acc}-\dot{P}_{b}^{\rm Shk},
\end{align}
where $\dot{P}_{b}^{\rm Acc}$ is caused by the difference of accelerations of
binary pulsars and the Solar System projected along the line of sight to the
pulsar;  $\dot{P}_{b}^{\rm Shk}$ is the ``Shklovskii''
effect~\cite{1970SvA....13..562S} caused by the relative kinematic motions of
the binary pulsar with respect to the Solar System barycenter.

\subsection{Individual bounds on the graviton mass}

From above descriptions, we can use the fractional discrepancy $\Delta$ of
binary pulsars to get an individual limit on the graviton mass {\it per}
pulsar.  We assume the measured discrepancy $\Delta$ to be normally distributed
about its unknown actual value~\cite{Finn:2001qi}.  The 1-$\sigma$ uncertainty
in $\Delta$ (see Table~\ref{tab:psr}) is of the same order as the central
value, so it must be properly accounted for in the analysis.  The squared mass
of graviton must be non-negative, but the standard confidence belts can not
guarantee to always give $\Delta$ a physical interval~\cite{Feldman:1997qc}.
As discussed above we choose an ordering principle to specify uniquely the
acceptance intervals which can avoid unphysical results. Using Table X in
Ref.~\cite{Feldman:1997qc}, we choose the $90\%$ C.L. confidence intervals
which are obtained by an ordering principle to replace the standard confidence
intervals.

We carefully choose  9 binary pulsars, and use the parameters in
Table~\ref{tab:psr} and Eq.~(\ref{eq:3}) to calculate the $90\%$ C.L. upper
limit on the graviton mass bound individually.  The relevant parameters of 9
binary pulsars are listed in Table~\ref{tab:psr}, and our results are listed in
Table~\ref{tab:massdata}.  The bounds in Table~\ref{tab:massdata} are sorted by
the strength in constraining $m_g$.  We notice that the updated parameters of
PSRs~B1913+16 and B1534+12 give tighter mass bounds than the widely used mass
bound in Ref.~\cite{Finn:2001qi}. This is because of longer observational
span and higher quality of timing data.  The best single limit on the graviton
mass comes from the Hulse-Taylor pulsar PSR B1913+16, which gives,
\begin{equation}
  m_{g}<3.5\times10^{-20}\,{\rm eV}/c^{2} \quad \mbox{(90\% C.L.)} \,. 
\end{equation}
It improves the previous limit from this pulsar (see Table I in
Ref.~\cite{Finn:2001qi}) by a factor of three.

\bgroup
\def\arraystretch{1.25}
\begin{table}
  \caption{\label{tab:massdata}
	The mass bounds from  individual binary pulsar systems at  $90\%$ C.L.,
      sorted by their constraints on $m_g$.}
  \begin{tabular}{p{4cm}p{4.5cm}}
  \hline\hline
    PSR \ \ \   & Graviton mass upper bound \\
    \hline
    B1913+16    &  $3.5\times10^{-20}\,{\rm eV}/c^{2}$\\
    J0737$-$3039  &  $4.3\times10^{-20}\,{\rm eV}/c^{2}$\\
    B1534+12    &  $5.2\times10^{-20}\,{\rm eV}/c^{2}$\\
    J1738+0333  &  $1.1\times10^{-19}\,{\rm eV}/c^{2}$\\
    J2222$-$0137  &  $1.8\times10^{-19}\,{\rm eV}/c^{2}$\\
    J1909$-$3744  &  $2.2\times10^{-19}\,{\rm eV}/c^{2}$\\
    J1012+5307  &  $2.9\times10^{-19}\,{\rm eV}/c^{2}$\\
    J1713+0747  &  $3.0\times10^{-19}\,{\rm eV}/c^{2}$\\
    J0348+0432  &  $6.0\times10^{-19}\,{\rm eV}/c^{2}$\\
    \hline
  \end{tabular}
\end{table}
\egroup

\subsection{A combined bound on the graviton mass}

 We use Bayesian statistics to combine the information from observations of
 several pulsar systems, and utilize the relevant parameters of 9 binary
 pulsars in Table~\ref{tab:psr} to give a combined bound on the graviton mass.
 There are no unknown parameters $\bm{\Xi}$ in this study, so we can rewrite
 Eq.~(\ref{eq:4}) to,
\begin{align}
   P\left( \left. m_g  \right| {\cal D}, {\cal H}, {\cal I} \right) &=
	\frac{P \left( \left.{\cal D}  \right| m_g, {\cal H}, {\cal
		I} \right) P\left( \left. m_g \right| {\cal H}, {\cal
		I} \right)}{P \left( \left. {\cal D} \right| {\cal H}, {\cal I}
		    \right)}  \,.
   \label{eq:5}
\end{align}
We assume that the observations of different binary pulsars are statistically
independent, so the combined likelihood can be split into the product of
multiple individual likelihoods,
\begin{align}
 P \left( \left. {\cal D} \right| m_{g}, {\cal H}, {\cal I} \right)=
   \prod^{n}_{i=1}P \left( \left. {\cal D}_{i} \right| m_{g}, {\cal H}, {\cal
     I} \right)\,,
\end{align}
where $n$ is the number of binary pulsars.  To be more explicit, we choose the
individual likelihood,
\begin{align}
 P \left( \left. {\cal D}_{i} \right| m_{g}, {\cal H}, {\cal I} \right) \propto
   {\rm exp} \left[ -\frac{1}{2}\frac{ \left(\dot{P}^{m_g}_{b}+\dot{P}^{\rm
   GR}_{b}-\dot{P}_{b}^{\rm intr} \right)^{2}}{\sigma^{2}} \right] \,,
\end{align}
where $\sigma$ is total uncertainty including the observational uncertainty and
the modeling uncertainty from $\dot P_b^{\rm Acc}$, $\dot P_b^{\rm Shk}$, and
so on; see Eq.~(\ref{eq:Pbdot:intr}); $\dot{P}^{m_g}_{b}$ is  the possible
contribution from a nonzero graviton mass,
\begin{align}
 \dot{P}^{m_g}_{b}\equiv\dot{P}_{b}-\dot{P}^{\rm GR}_{b} = \frac{5}{24}
 m_g^{2}\left(\frac{c^{2}P_{b}}{2\pi\hbar}\right)^{2}\frac{1}{F(e)}\dot{P}^{\rm
 GR}_{b}\,.
\end{align}

For the prior probability of the graviton mass, we use two un-informative
choices in this paper. One is to choose a uniform prior probability in $m_g$,
and the other is to choose a uniform prior probability in $\ln m_{g}$.  For
both prior probabilities, we take the graviton mass in the range  $m_{g}\in
\left[10^{-26},10^{-16} \right]\,{\rm eV}/c^{2}$. For ease in comparison, the
range of prior is the same as that chosen by the LIGO/Virgo
Collaboration~\cite{TheLIGOScientific:2016src}.

In Eq.~(\ref{eq:5}), we use the relevant parameters of binary pulsars in
Table~\ref{tab:psr} and our two different prior distributions to give two
different posterior distributions. The posteriors are shown in
Fig.~\ref{fig:post}. Because the prior uniform in $\ln m_g$ gives more support
to a smaller mass, overall it gives a better constraint in $m_g$. The apparent
drop in the posterior distribution, when the prior uniform in  $m_g$ is used,
is artificial in Fig.~\ref{fig:post} due to the usage of a logarithmic scale in
the abscissa axis.

We calculate the upper limit on the graviton mass from two posterior distributions. 
When the prior is uniform in $m_g$, we have,
\begin{equation}
  m_{g}<3.1\times10^{-20}\,{\rm eV}/c^{2}  \quad \mbox{(90\% C.L.)} \,.
\end{equation}
As one can see, this result is very close to the limit from the Hulse-Taylor
pulsar PSR~B1913+16 in the frequentist approach (see Table~\ref{tab:massdata}).

When the prior is uniform in $\ln m_g$, we have,
\begin{align}
  m_{g}<5.2\times10^{-21}\,{\rm eV}/c^{2} \quad \mbox{(90\% C.L.)}\,. 
  \label{eq:limit:ln}
\end{align}
This limit, though not as tight as some other limits~\cite{deRham:2016nuf},
improves the Finn-Sutton limit in 2002~\cite{Finn:2001qi} by a factor of more
than 10.

\begin{figure}
\includegraphics[width=0.40\textwidth]{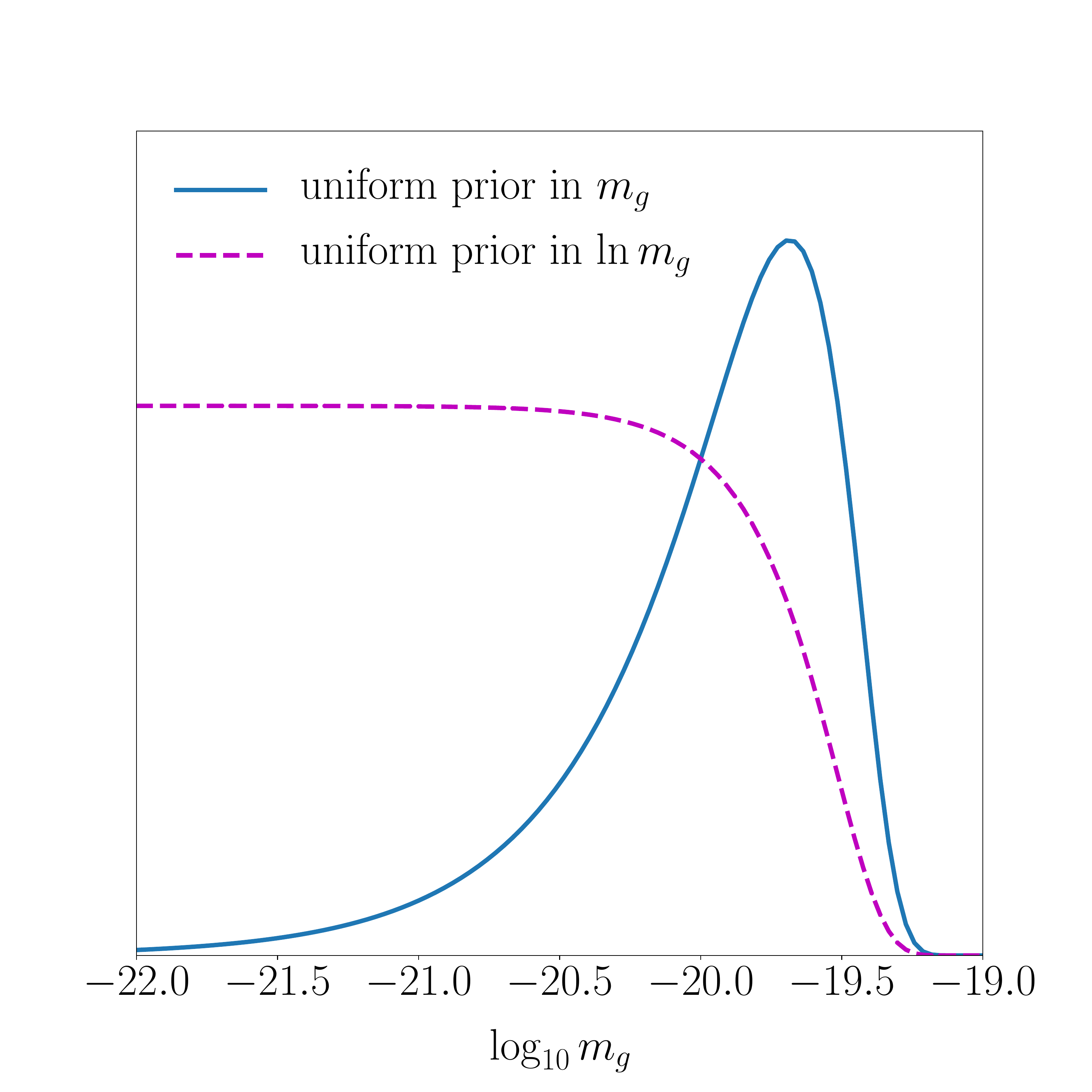}\\
\caption{The posterior distributions with two different priors for
$m_g$.}\label{fig:post}
\end{figure}

\section{Discussions}

During recent decades, because of the tremendous advance in constructing
theories for massive gravity~\cite{Dvali:2000rv, deRham:2010ik, deRham:2010kj,
Hassan:2011zd, deRham:2014zqa, deRham:2016nuf}, there are various means to give
graviton mass bounds, which generally can be classified into three
categories~\cite{deRham:2016nuf}.  We briefly list them below, and compare them
with our results in the following.
\begin{enumerate}
\item If the graviton is a massive boson, the gravitational force typically
  acquires an exponential Yukawa suppression, and the static gravitational
  potential is modified to $\Phi\propto e^{-m_{g}r}/r$. It is different from
  the case of massless gauge bosons where the potential has a $1/r$
  falloff~\cite{Will:2018gku}.
\item If gravity is propagated by a massive field, the massive graviton
  propagator will lead to a modified dispersion relation $E^2 = p^2 c^2 + m_g^2
  c^4$.  The direct result is that GWs no longer travel at the velocity of
  light but depend on their frequencies~\cite{Will:1997bb}.
\item For a spin-2 massive boson, there may be five degrees of freedom in the
  propagation, including two tensor modes, two vector modes, and one scalar
  mode. The extra modes could lead to a fifth force~\cite{deRham:2016nuf}.
\end{enumerate}
The above three arguments are not always true in specific massive gravity
theories, {\it e.g.} in the DGP model~\cite{Dvali:2000rv} and the dRGT
theory~\cite{deRham:2010kj} where screening mechanisms take
effect~\cite{Vainshtein:1972sx}. Nevertheless, they are effective in
communicating information in the community as simple phenomenological
treatments.

Our limit obtained in this paper is also a phenomenological one based on the
phenomenological action~(\ref{eq:action})~\cite{Visser:1997hd, Finn:2001qi}.
Below, we compare our result with those existing in the literature.

For Yukawa potential, the falloff is obvious when the distance from the source
is greater than the Compton wavelength of graviton, $\lambda_{g} \equiv h/m_g
c$. The most recent mass bound from ephemeris observations of the Solar System
is $m_{g}<(6-10)\times10^{-24}\,{\rm eV}/c^{2}$~\cite{Will:2018gku}.  This
result is three orders of magnitude better than our result in
Eq.~(\ref{eq:limit:ln}). However, our mass bound comes from a dynamical process
and can reflect the dynamics of binaries.  The Yukawa potential describes a
static field, and though the result from Yukawa potential is more competitive,
it can not reflect two-body dynamics.

Modified dispersion relation is based on the linearized theory of gravity,
which is similar to the Yukawa potential.  The frequency of GWs is increasing
during the binary inspiral.  If graviton is massive, according to its modified
dispersion relation the velocity of graviton should depend on the GW
frequency~\cite{Will:1997bb}.  It means that the lower-frequency GWs, which are
emitted earlier, propagate slower than the higher-frequency GWs which are
emitted later.  The different propagating velocities would distort the shape of
the observed GW waveforms~\cite{Will:1997bb}.  So one can get the mass bounds
from the GW detection, as were done by the LIGO/Virgo
Collaboration~\cite{TheLIGOScientific:2016src, Abbott:2017vtc,
LIGOScientific:2019fpa} (see Sec.~\ref{sec:intro}).  Compared with our result,
the limit from GWs is more competitive.  However, the mass bound coming from
the direct detection of GWs is an effect of accumulation in GW phases. It is of
different nature to the limit from binary pulsars which reflects a dynamical
process of a binary motion. These two kinds of limits are of a kinematic origin
for GWs and of a dynamic origin for binary pulsars.

For the fifth force, the bounds on the graviton mass come from the additional
modes.  In general, studies focus on the scalar mode and neglect the vector
modes. Although the vector modes can induce a fifth force in principle, they
usually do not couple to matter in the decoupling limit~\cite{deRham:2016nuf}.
The Finn-Sutton method focuses on the two transverse tensor modes only, which is
different from the fifth force.  The mass bounds from the fifth force are
tightest, such as the ones from the lunar laser ranging experiments in the DGP
model~\cite{Dvali:2000rv} which is $m_{g}<10^{-32}\,{\rm
eV}/c^{2}$~\cite{Dvali:2002vf}.  But these bounds are model dependent, such as
for the DGP model and the dRGT theory~\cite{deRham:2010ik,deRham:2010kj}.  In
comparison, we consider our limit much more model-independent.

Our limit in Eq.~(\ref{eq:limit:ln}) from  binary pulsars is not as tight as
the other results, but it is the only dynamical  bound which comes from binary
pulsars (see however, Refs.~\cite{deRham:2012fw, deRham:2012fg}). We consider
it {\it complementary} to bounds from other means (also {\it complementary} to
the limits from binary pulsars in a model-specific treatment).  In the future,
we can improve the binary pulsar measurements and achieve smaller uncertainties
on the timing parameters with the Five-hundred-meter Aperture Spherical
Telescope (FAST) and the Square Kilometre Array~\cite{Kramer:2004hd,
Shao:2014wja, Bull:2018lat}.  Especially, the observed timing precision on the
$\dot P_b$ is proportional to $T^{-5/2}$ where $T$ is the observational span.
Therefore, if we can model the Milky Way accurately to account for the extra
contributions in $\dot P_b^{\rm obs}$ [see Eq. (\ref{eq:Pbdot:intr})], the
limit from the Finn-Sutton method improves quickly.  Then we will get a tighter
mass bound on graviton in a dynamic regime with binary pulsars. 

Finally, as was pointed out in Sec.~\ref{sec:graviton}, we would like
  to stress that, same as the well-recognized limit from \citet{Finn:2001qi},
  our limits are based on the phenomenological action (\ref{eq:action}). It is
  a phenomenological treatment of a linearized version of a massive rank-2
  tensor field $g_{\mu\nu}$; it should not be taken as a full and
  sophisticatedly designed theory.  The behavior of the tensor modes is the
  same as that in a healthy theory of massive gravity and the bound which comes
  from the Finn-Sutton method is in principle applicable to binary pulsar
  systems~\cite{deRham:2016nuf}.  This viewpoint also applies to the limits
  obtained from the propagation of GWs and the Yukawa potential. These {\it
  generic} limits do not necessarily mean that they are applicable to all
  theories of massive gravity~\cite{deRham:2014zqa, deRham:2016nuf}.
Nevertheless, they are still useful as simple empirical results on the mass of
graviton. 

\acknowledgments

We are grateful to Yi-Fu Cai and Norbert Wex for discussions, John Antoniadis
for reading the manuscript, and the anonymous referees for useful
comments.  This work was supported by the Young
Elite Scientists Sponsorship Program by the China Association for Science and
Technology (2018QNRC001).  It was partially supported by the National Natural
Science Foundation of China (11721303, 11475006), the Strategic Priority
Research Program of the Chinese Academy of Sciences through the grant No.
XDB23010200, and the European Research Council (ERC) for the ERC Synergy Grant
BlackHoleCam under Contract No.  610058.


%

\end{document}